\documentstyle{article}
\begin{document}
\LARGE
\begin{center}
\bf Creation of Closed or Open Universe from Constrained
Instanton
\vspace*{0.4in}
\normalsize \large \rm 

Z.C. Wu

Dept. of Physics

Beijing Normal University

Beijing 100875, China

\vspace*{0.6in}
\large
\bf
Abstract
\end{center}
\vspace*{.1in}
\rm
\normalsize

In the no-boundary universe the universe is created from an
instanton. However, there does not exist any instanton for the
``realistic'' $FRW$  universe with a scalar field. The
``instanton'' leading to its quantum creation may be modified and
reinterpreted as a constrained gravitational instanton.

\vspace*{0.8in}

PACS number(s): 98.80.Hw, 98.80.Bp, 04.60.Kz, 04.70.Dy


\vspace*{0.8in}

e-mail: wu@axp3g9.icra.it $\;\;\;\;$

\pagebreak

A gravitational instanton is defined as a Euclidean solution to
the Einstein field equation. In the no-boundary universe it is
believed that the universe is created through a quantum
transition at its equator to the Lorentzian spacetime. At the
equator, all canonical momenta should vanish.

However, there exist too few instantons for a vacuum model with a
positive cosmological constant $\Lambda$ [1]. Even for a model
with matter fields, if the situation is not worse, it is
not expected to be improved dramatically. 

In quantum cosmology, the scalar model has been extensively
investigated [2]. It is a closed $FRW$ universe coupled to a
scalar field $\phi= \phi (t)$ with potential $V(\phi)$. Its
Euclidean metric is described by
\begin{equation}
ds^2 =  d\tau^2 + b^2(\tau)(d\chi^2 + \sin^2\chi (d\theta^2 +
\sin^2\theta d\psi^2)).
\end{equation}
The field $\phi$ obeys the equation 
\begin{equation}
\ddot{\phi} + 3 \frac{\dot{b}}{b}\dot{\phi} = V_{,\phi},
\end{equation}
where dots denote derivatives with respect to the imaginary time.
The model is called the Hawking model if the potential takes the
form  $V(\phi) = m^2\phi^2$, where $m$ is the mass of the scalar.
For a
general case the potential is chosen such that the evolution of
the spacetime and the scalar can be described as follows [4]: In
the Euclidean regime and the Planckian inflationary era, the
model can be approximated by a de Sitter model. The derivative of
the scalar with respect to the imaginary time should be zero at
the south pole of the Euclidean solution. This is required by the
regularity condition imposed there by the no-boundary proposal
[3]. In this model there is an effective cosmological constant
parametrized by the initial value $\phi_0$ of the scalar field at
the south pole. To make the model more realistic, one has to
assume that $V_{, \phi}(\phi_0)\neq 0$. 

In the Euclidean regime, the scalar will increase slowly
before the universe reaches its maximum size. However, as soon as
the universe begin to contract, it increases rapidly and
eventually diverges logarithmically as the universe collapses to
a singularity. It was argued that the singular behavior is
precisely critical to avoid a pathology in quantum mechanics [4].
The Euclidean action for the solution is [4]
\begin{equation}
\bar{I} \approx - \frac{12\pi^2 M^4_{Pl}}{V(\phi_0)},
\end{equation}
where  $M_{Pl}=(8\pi G)^{-1/2}$ is the reduced Planck mass. We
assume that $d\bar{I}/dV(\phi_0) \neq 0$ over the whole range of
$\phi_0$. We call this model as the ``realistic'' model, and the
Hawking model becomes one of its special cases.

If the potential is a constant or the slope $V_{, \phi}(\phi_0)$
of the potential at the initial value $\phi_0$ is zero, then its
effect is exactly the same as a cosmological constant and the
model is identical to the de Sitter model, and the scalar field
will remain a constant. This is because the field $\phi$ obeys
Eq. (2), and the derivative of the scalar with respect to the
imaginary time should be zero at the south pole of the manifold.
This is not a realistic model.

For the ``realistic'' model, the condition $V_{,
\phi}(\phi_0)\neq
0$  will lead to 
\begin{equation}
\frac{d\bar{I}}{d\phi_0} \neq 0,
\end{equation}
which means that the action is not stationary, at least with
respect to the variation of the initial scalar field $\phi_0$.
Therefore the distorted 4-sphere does not qualify as an instanton
in the ordinary sense [5].

In fact people have long realized that for the Hawking
model [1] there is no regular Euclidean solution. Or
equivalently, one
cannot find a Euclidean regular solution with a 3-geometry (the
equator) as the only boundary on which the second fundamental
form vanishes and the normal derivative of the matter field is
zero. At best, one can only find a Euclidean
solution with approximately vanishing momenta at the equator [6].
Indeed, for a general model the following four conditions are
equivalent [7][8]: (i)
The Euclidean manifold satisfies the field equation everywhere,
(ii) its Euclidean action is stationary, (iii) there is no
singularity in the Euclidean solution, and (iv) the Euclidean
solution has a boundary, the equator, at which the canonical
momenta vanish.  

One way out of the trouble caused by singularity behavior in the
scalar model is to reinterpret the
Euclidean  solution to the field equation as a
constrained gravitational instanton [8]. The south hemisphere of
the
manifold is the stationary action solution under the condition
that, at the maximum size where the quantum
transition is supposed to occur, the 3-geometry is given. The
whole manifold is
made by joining this south hemisphere and its oriented reversal
as the north hemisphere. One can use $\phi_0$ to parametrize the
3-geometry, and then Eq. (4) will no longer bother quantum
cosmologists. The variational
calculation shows that the stationary action solution should be
regular and satisfy the field equations everywhere,  with the
only
possible exception at the 3-geometry equator. Therefore the joint
manifold has a stationary action under the restriction imposed at
the equator or for a fixed $\phi_0$ and qualifies as a
constrained gravitational instanton, which can be used as the
seed for the creation of the universe. For this model, the
canonical momentum of the scalar field at the equator, i.e, its
normal derivative, is allowed to be
nonzero. One may try to find a complex solution as a
seed for the creation of the universe. The complex solution may
offer some help to obtain a real evolution in the Lorentzian
regime [6].

One may try a model with a Euclidean action form
qualitatively different from Eq. (3). Then there may exist some
discrete values of $\phi_0$ for which the actions are stationary.
Then we obtain true instantons [9]. However, it will not
occur for a simple case like the Hawking model and the
``realistic'' model with the action approximated by Eq. (3).

We can even offer a stronger argument for this point [8].
In general, if the wave function of the universe represents an
ensemble of evolutions with continuous parameters, like the case
of this model, then it is unlikely that all these trajectories of
the ensemble can be obtained through analytic continuations
from an ensemble of instantons. Unless the action is a constant
function of the parameters, then the action cannot be stationary
with respect to them over their range. On the other hand, the
condition for the existence of an instanton in the ordinary sense
is that its action must be stationary. As far as the singularity
problem or the problem of nonstationary action is concerned, the
complex solutions offer no help to the Hawking model.

The constrained gravitational instanton has a nice application to
the model of a single black hole creation in the de Sitter
background [8]. The 3-geometry of the equator is parametrized by
the
mass , charge and angular momentum parameters of the black hole
created. The Euclidean action is stationary only under the
restriction of the given 3-geometry. Therefore the corresponding
Euclidean solution is a constrained gravitational instanton. If
one lifts the constraints at the equator, one can only get two
regular
gravitational instantons, the $S^4$ and $S^2 \times S^2$ spaces.
They are the seeds for the de Sitter universe with no black hole
and Nariai universe with two black holes. These two possibilities
can be contrasted with the single black hole case. The $S^4$
solution yields the maximum probability of black hole creation,
$S^2\times S^2$ yields the minimum creation probability.

The $S^4$ space possessing $O(5)$ invariance can be analytically
continued into de Sitter space with 3-spheres as spatial sections
or continued into anti-de Sitter space with 3-hyperboloids as
spatial sections [10]. In the ``realistic'' model, the distorted
4-sphere
can also be continued into an open $FRW$ universe [4]. As long as
we are working in the minisuperspace model, everything is fine.
It is noted that the $O(4)$ invariance of the distorted 4-sphere
is crucial to the continuation. 
Now the great task quantum cosmologists have to face is, how to
carry out the fluctuation calculation on the open background?
This kind of calculation is crucial in predicting the structure
of the
universe. The origin of this difficulty is that we lack a general
proposal for the quantum ground state of an open universe. To
make the open model work, one has
to overcome this great obstacle.

\vspace*{0.3in}

\bf References:

\vspace*{0.1in}
\rm

1. G.W. Gibbons and J.B. Hartle, \it Phys. Rev. \rm \bf D\rm
\underline{42}, 2458 (1990).

2. S.W. Hawking, \it Nucl. Phys. \rm \bf B\rm \underline{239},
257 (1984).

3. J.B. Hartle and S.W. Hawking, \it Phys. Rev. \rm \bf D\rm
\underline{28}, 2960 (1983).

4. S.W. Hawking and N. Turok, hep-th/9802030; gr-qc/9802062. For
related ideas, see A. Linde, gr-qc/9802038.

5. For related ideas, see A. Vilenkin, hep-th/9803084.

6. R. Bousso and S.W. Hawking, \it Phys. Rev. \rm \bf D\rm
\underline{52}, 5659 (1995).

7. X.M. Hu and Z.C. Wu, \it Phys. Lett. \rm \bf B\rm
\underline{149}, 87 (1984).

8. Z.C. Wu, \it Int. J. Mod. Phys. \rm \bf D\rm\underline{6}, 199
(1997); \it Gene. Rela. Grav. \rm\underline{30}, 115 (1998); gr-
qc/9712066.

9. L. Jensen and P.J. Steinhardt, \it Nucl. Phys.  \rm \bf B\rm
\underline{237}, 176 (1984).

10. Z.C. Wu, \it Phys. Rev. \rm \bf D\rm
\underline{31}, 3079 (1985).

\end{document}